\documentclass[lettersize,journal,comsoc]{IEEEtran}
\usepackage{amsfonts}
\usepackage{amssymb}
\usepackage{eurosym}
\usepackage{cite}
\usepackage{graphicx}
\usepackage{epstopdf}
\usepackage{amsmath}
\usepackage{tikz,lipsum}
\usepackage{caption}
\usepackage[T1]{fontenc}
\usepackage{amsthm}
\usepackage{mathrsfs}
\usepackage{color}
\usepackage{stfloats}
\usepackage{xcolor, etoolbox}
\usepackage{subcaption}
\usepackage{comment}
\usepackage{algcompatible}
\usepackage[ruled]{algorithm2e}

\def\baselinestretch{0.95}
\IEEEoverridecommandlockouts

\newtheorem{remark}{Remark}

\begin{document}

\title{Secure Pinching Antenna-aided ISAC}
\author{Elmehdi Illi, \IEEEmembership{Member, IEEE}, Marwa Qaraqe, %
\IEEEmembership{Senior Member, IEEE}, and Ali Ghrayeb, %
\IEEEmembership{Fellow, IEEE} \thanks{%
The authors are with the College of Science and Engineering, Hamad Bin
Khalifa University, Qatar Foundation, Doha, Qatar. (e-mails: \{eilli,
mqaraqe, aghrayeb\}@hbku.edu.qa).} }
\maketitle

\begin{abstract}
In this letter, a pinching antenna (PA)-aided scheme for establishing a
secure integrated sensing and communication system (ISAC) is investigated.
The underlying system comprises a dual-functional radar communication (DFRC)
base station (BS) linked to multiple waveguides to serve several downlink
users while sensing a set of malicious targets in a given area.\ The
PA-aided BS\ aims at preserving communication confidentiality with the
legitimate users while being able to detect malicious targets. One objective
of the proposed scheme is to optimize the PA locations, based on which an
optimal design of the legitimate signal beamforming and artificial noise
covariance matrices is provided to maximize the network's sensing
performance, subject to secrecy and total power constraints. We demonstrate
the efficacy of the proposed scheme through numerical examples and compare
that against a traditional DFRC ISAC system with a uniform linear array of
half-wavelength-spaced antennas. We show that the proposed scheme
outperforms the baseline PA-aided scheme with equidistant PAs by $3$ dB in terms of
illumination power, while it can provide gains of up to $30$ dB of the same
metric against a traditional ISAC system with half-wavelength-space uniform
linear arrays.
\end{abstract}


\section{Introduction}

Integrated sensing and communication (ISAC) has emerged in the past few
years as a means of supporting power- and spectrum-efficient sensing
applications by utilizing the same hardware, frequency, and power resources
for both sensing and communication tasks \cite{mmimoisac}. As well, the
multiple-input multiple-output (MIMO) technology has been forming a key
pillar in the fifth generation of cellular networks (5G), due to its great
potential in enabling received signal strength enhancement and spatial
multiplexing \cite{mmimo}. MIMO demonstrated notable gains in establishing
robust communications, sensing, and ISAC schemes \cite{mmimoisac}.
Nevertheless, MIMO can hardly turn an unfavorable channel to an advantageous
one. To this end, movable antenna systems (MAS) and reconfigurable
intelligent surfaces (RIS) techniques have been proposed as supportive
techniques to MIMO. While MAS is based on adjusting antenna positions by few
wavelengths, RIS is based on manipulating signal reflections through
optimized phase shifts of its reflective elements. However, due to antenna
position's tunability over only a few wavelengths, FAS fails in solving one
of the MIMO main issues, that is, the line-of-sight (LoS) path blockage,
while RIS suffers from the double path-loss, especially at higher
frequencies.

A promising alternative, referred to as pinching-antenna systems (PASS), has
been recently introduced by NTT DOCOMO \cite{docomo}. PASS uses low-loss
long dielectric waveguides to guide signals and radiate them through small
dielectric elements, also known as pinching antennas (PAs) manually placed
along the waveguide in positions of interest. This setup exhibits several
benefits compared to traditional MIMO, FAS, and RIS in overcoming LoS
blockages and high free-space path loss (FSPL) in mmWave- and THz-band
transmissions, and broadening network coverage in indoor and outdoor
scenarios, thanks to its flexible antenna placement. Such a
propagation-tuning feature makes PASS well suited for establishing physical
layer security (PLS), where beamforming can be optimized to strengthen
communication with legitimate users while weakening it with eavesdroppers.

The past year witnessed a notable growth in the number of works that
designed and analyzed PASS in various setups. For instance, \cite{lit1}
analyzed the ergodic capacity of PASS. The work in \cite%
{lit2,lit3,lit4,lit7} proposed an optimization framework for PASS by
optimizing transmit PA locations, beamforming vector, and uplink transmit
power to either maximize\ (minimize) the sum rate\ (transmit power) subject
to power (sum rate) constraints. The authors of \cite{lit11} maximized the
network secrecy capacity (SC)\ with the use of an optimized artificial noise
(AN), transmit beamforming, and PA positions. In \cite%
{isaclit1,isaclit2,isaclit4}, the authors analyzed the sensing and
reliability performance of PA-aided ISAC systems (PA-ISACS), showing the
potential of PASS to establish robust sensing and reliable communication in
various scenarios.

Despite the aforementioned contributions, most of them were limited to
maximizing the network reliability in terms of its achievable rate, sum
rate, or secrecy rate. Furthermore, the work in \cite%
{isaclit1,isaclit2,isaclit4} focused on PA-ISACS analysis and/or
optimization only from the perspective of sensing and communication
reliability, while missing the inclusion of the security aspect. Motivated
by the above, this work aims to analyze and optimize the secrecy and sensing
performance of a secure PA-ISACS. Using a suboptimal designed PA positions
optimization scheme, along with semi-definite programming for optimal
beamforming and AN covariance, a robust optimization framework is proposed
to maximize the per-target sensing illumination power, subject to secrecy
and total power constraints. The proposed scheme exhibits at least $3$-
and $30$-dB gains in terms of sensing illumination power against a baseline
PA-ISACS with equispaced PAs and against an ISAC system with a uniform
linear array (ULA) of half-wavelength-spaced antennas, respectively.


\section{System Model}

\begin{figure}[b]
\par
\vspace*{-.3cm}\begin{center}
\includegraphics[scale=.30]{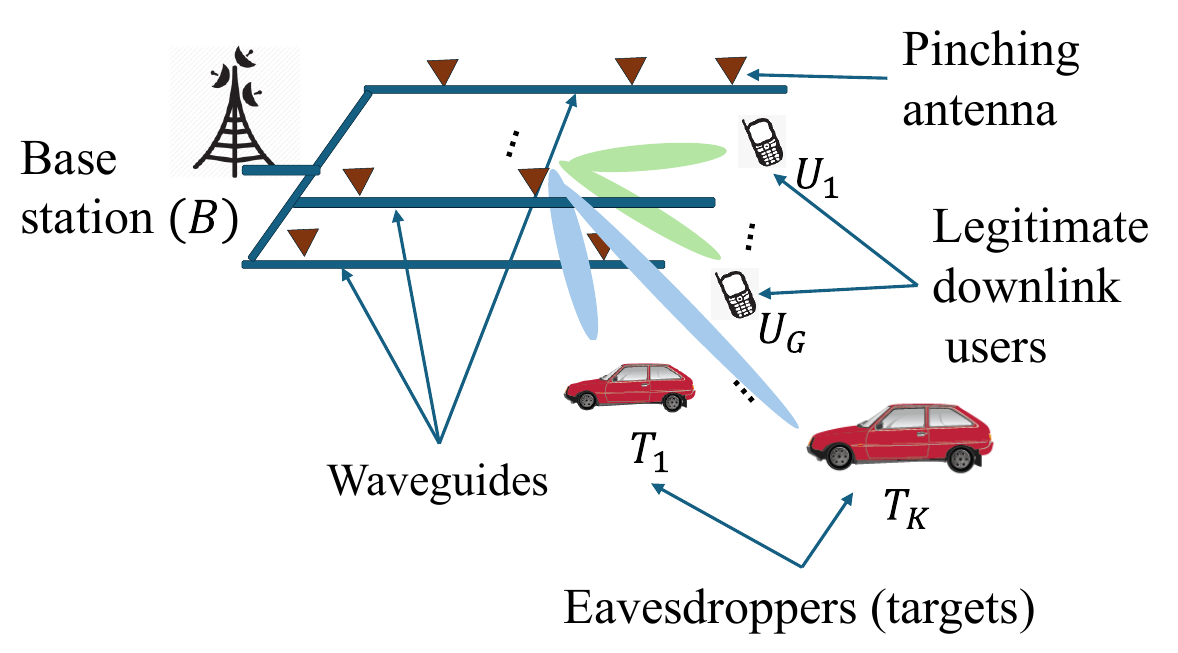}
\end{center}
\par
\captionsetup{font=footnotesize}
\caption{Considered pinching-antenna-enabled ISAC system.}
\label{sysmod}
\end{figure}

\subsection{Signal and Channel Model}

Consider a PA-ISACS system, as given in Fig. \ref{sysmod}, comprising a base station (BS), labeled $B$,
communicating with a set of $G$ users $\{ U_{g}\} _{g=1}^{G}$.
Furthermore, a set of $K$ malicious targets $\left\{ T_{k}\right\} _{k=1}^{K}
$ are present, where $B$, acting as an ISAC transceiver aims at sensing and
detecting their presence in the pre-known locations. Furthermore, $\left\{
T_{k}\right\} _{k=1}^{K}$ aim to compromise transmission by eavesdropping on
legitimate signals of the $G$ users. $B$ is connected to $N$ parallel
dielectric waveguides through flexible cables. It is assumed that the $n$th
waveguide is deployed in parallel to the $x$ axis and at given $y$ and $z$
coordinates, denoted by $y_{n}$ and $z_{n}$. To this end, $M_{t}$ transmit
PAs and $M_{r}$ receive ones can be activated along each waveguide to
enhance (i)\ the transmission channel to the set of users and targets and
(ii) echo signals reception from the $K$ targets. We denote by $(
x_{n,m},y_{n,m},z_{n,m}) $ the Cartesian coordinates of the $m$th
transmit PA\ on the $n$th waveguide. Notably, due to the deployment of the
parallel waveguides along the $x$ axis, it yields $y_{n,m}=y_{n}$ and $%
z_{n,m}=z_{n}$ $\forall m=1,\ldots ,M_{t}$. In addition, the coordinates for
each user $U_{g}$ and target $T_{k}$ are denoted as $(
x_{U_{g}},y_{U_{g}},z_{U_{g}}) $ and $(
x_{T_{k}},y_{T_{k}},z_{T_{k}}) $, respectively. The received signal at
$U_{g}$ and $T_{k}$ can be expressed as \cite{lit1}%
\begin{equation}
y_{\Lambda }=\mathbf{g}_{B\Lambda }\left( \mathbf{x,y,z}\right) \mathbf{H}%
\left( \mathbf{x}\right) \mathbf{s}+w_{\Lambda },\Lambda \in \left\{
U_{g},T_{k}\right\}   \label{rxsig}
\end{equation}%
where $\mathbf{g}_{B\Lambda }( \mathbf{x}) \triangleq $ $[
\mathbf{g}_{B_{1}\Lambda }\left( \mathbf{x}_{1},y_{1},%
z_{1}\right) ,\ldots ,\mathbf{g}_{B_{N}\Lambda }\left( \mathbf{x}_{N},y_{N},z_{N}\right) ] \in
\mathbb{C}
^{1\times M_{t}N}$ is the channel response of the link between all the
network's PAs and node $\Lambda $, and
\begin{equation}
\mathbf{g}_{B_{n}\Lambda }\left( \mathbf{x}_{n},y_{n},z_{n}\right) =\left[
\begin{array}{c}
\sqrt{\zeta _{B_{n,1}\Lambda }}e^{-j2\pi /\lambda d_{B_{n,1}\Lambda
}},\ldots , \\
\sqrt{\zeta _{B_{n,M_{t}}\Lambda }}e^{-j2\pi /\lambda d_{B_{n,M_{t}}\Lambda
}}%
\end{array}%
\right] \in
\mathbb{C}
^{1\times M_{t}}  \label{glos}
\end{equation}%
represents the channel response vector between the $n$th waveguide and node $%
\Lambda $. Additionally,
\begin{equation}
\zeta _{B_{n,m}\Lambda }=G^{(B_{n,m})}G^{(\Lambda )}\left( \frac{\lambda }{%
4\pi d_{B_{n,m}\Lambda }}\right) ^{2}  \label{fspl}
\end{equation}%
denotes the FSPL term between the $m$th antenna of the $n$th waveguide,
labeled $B_{n,m}$, and $\Lambda $, and $G^{(B_{n,m})}$ is the transmit gain
of $B_{n,m}$. Also, $G^{(\Lambda )}$ is node $\Lambda $'s receive antenna
gain, $\lambda $ is the signal wavelength, and
\begin{equation}
d_{B_{n,m}\Lambda }=\sqrt{\left( x_{n,m}-x_{\Lambda }\right) ^{2}+\left(
y_{n}-y_{\Lambda }\right) ^{2}+\left( z_{n}-z_{\Lambda }\right) ^{2}}
\label{distance}
\end{equation}%
is the distance between $B_{n,m}$ and node $\Lambda $. Also, $\mathbf{x}_{n}=%
\left[ x_{n,1},\ldots ,x_{n,M_{t}}\right] $ denotes the position vector for
the $M_{t}$ antennas of the $n$th waveguide along the $x$-axis with $\mathbf{%
x}\triangleq \left[ \mathbf{x}_{1},\ldots ,\mathbf{x}_{N}\right] $, while $%
\mathbf{y\triangleq }\left[ y_{1},\ldots ,y_{N}\right] $ and $\mathbf{%
z\triangleq }\left[ z_{1},\ldots ,z_{N}\right] $ are the waveguides
positions along the $y$ and $z$ axes, respectively.

\begin{remark}
By assuming that the waveguides are deployed over fixed $y_{n}$ and $z_{n}$
values, the channel vectors' dependence on $y_{n}$ and $z_{n}$ will be
omitted in the sequel and the considered PA-ISACS performance evaluation and
optimization will be performed with respect to the PA positions over the $x$%
-axis.
\end{remark}

On the other hand, we define%
\begin{equation}
\mathbf{H}\left( \mathbf{x}\right) \triangleq \left[
\begin{array}{cccc}
\mathbf{h}\left( \mathbf{x}_{1}\right)  & \mathbf{0}_{M_{t}\times 1} &
\ldots  & \mathbf{0}_{M_{t}\times 1} \\
\mathbf{0}_{M_{t}\times 1} & \mathbf{h}\left( \mathbf{x}_{2}\right)  &
\ldots  & \mathbf{0}_{M_{t}\times 1} \\
\vdots  & \vdots  & \ddots  & \vdots  \\
\mathbf{0}_{M_{t}\times 1} & \mathbf{0}_{M_{t}\times 1} & \ldots  & \mathbf{h%
}\left( \mathbf{x}_{N}\right)
\end{array}%
\right] \in
\mathbb{C}
^{NM_{t}\times N}  \label{Hx}
\end{equation}%
as the overall in-waveguide channel matrix for the $N$ waveguides with $%
\mathbf{0}_{M_{t}\times 1}$ being an all-zeros column vector of $M_{t}$
elements, and%
\begin{equation}
\mathbf{h}\left( \mathbf{x}_{n}\right) =\left[
\begin{array}{c}
\sqrt{\theta _{n,1}}e^{-j2\pi /\lambda n_{g}x_{n,1}},\ldots  \\
,\sqrt{\theta _{n,M_{t}}}e^{-j2\pi /\lambda n_{g}x_{n,M_{t}}}%
\end{array}%
\right] ^{T}\in
\mathbb{C}
^{M_{t}\times 1}  \label{hx}
\end{equation}%
as the in-waveguide propagation vector for the $n$th waveguide, which
represents the signal attenuation from $B$ to $B_{n,m}$, and $\theta _{n,m}$
is the ratio of signal power radiated by $B_{n,m}$. Note from (\ref{rxsig})
that each waveguide's antennas radiate the same signal, increasing the
transmit power level. Without loss of generality, an equal-power model is
considered as utilized in \cite[Eqs. (20), (21)]{lit1}, i.e., $\theta
_{n,m}=\theta ,\forall n,m$, where $0<\theta \leq 1/M_{t}$. Also, $n_{g}$
defines the refractive index of the waveguide medium, and $\mathbf{s=Vu+w}%
\in
\mathbb{C}
^{N\times 1}$is the signal vector containing the $N$ transmit signals by the
$N$ waveguides. Furthermore, $\mathbf{V=}\left[ \mathbf{v}_{1},\ldots ,%
\mathbf{v}_{G}\right] \in
\mathbb{C}
^{N\times G}$ is the pinching beamforming matrix used to beamsteer the $G$
users' signals simultaneously by the PASS composed of $M_{t}N$ radiating
antennas, where $\mathbf{v}_{g}\in
\mathbb{C}
^{N\times 1}$ denotes $U_{g}$'s beamforming vector, and $\mathbf{u}=\left[ {u%
}_{1},\ldots ,{u}_{G}\right] \in
\mathbb{C}
^{G\times 1}$ is the raw data signal vector of the $G$ users. On the other
hand, the system beamforms an AN signal $\mathbf{w}\in
\mathbb{C}
^{N\times 1}$ with the objective of (i)\ degrading the decoding performance
at the malicious targets and increase the degrees of freedom in sensing
their presence. $\mathbf{w}$ is considered as a zero-mean complex Gaussian
vector with covariance matrix $\mathbf{W=}\mathbb{E}\left[ \mathbf{ww}^{H}%
\right] $, where $(.)^{H}$ dentoes the Hermitian of a vector/matrix and $%
\mathbb{E}\left[ \mathbf{.}\right] $ is the expectation operator. Both the
legitimate signal and AN\ beamforming are subject to a total power budget
constraint, i.e., $\sum_{g=1}^{G}\mathrm{Tr}\left[ \mathbf{V}_{g}%
\right] +\mathrm{Tr}\left[ \mathbf{W}\right] \leq $ $P_{\max }$, where $%
\mathbf{V}_{g}\triangleq \mathbf{v}_{g}\mathbf{v}_{g}^{H}$, $P_{\max }$ is
the system power budget, and $\mathrm{Tr}\left[ \mathbf{.}\right] $ is the
trace of a matrix. Lastly, $w_{\Lambda }$ is the zero-mean additive white
Gaussian noise (AWGN) at node $\Lambda $ of variance $\sigma _{\Lambda }^{2}$%
. Per the received signal formula in (\ref{rxsig}), and exploiting the
properties of positive semidefinite (PSD) matrices, the received
signal-to-interference-and-noise ratio (SINR) at $U_{g}$ and $T_{k}$ to
decode $u_{g}$ can be expressed as
\begin{equation}
\gamma _{\Lambda }^{(g)}\left( \left\{ \mathbf{V}_{g}\right\} _{g=1}^{G},%
\mathbf{W,x}\right) =\frac{\mathrm{Tr}\left( \mathbf{F}_{B\Lambda }\left(
\mathbf{x}\right) \mathbf{V}_{g}\right) }{\mathrm{Tr}\left( \mathbf{F}%
_{B\Lambda }\left( \mathbf{x}\right) \left[ \mathbf{W+}\sum\limits
_{\substack{ g^{\prime }=1 \\ g^{\prime }\neq g}}^{G}\mathbf{V}_{g^{\prime }}%
\right] \right) +\sigma _{\Lambda }^{2}},  \label{snruser}
\end{equation}
where $\mathbf{F}_{BT_{k}}( \mathbf{x}) \triangleq \mathbf{f}%
_{B\Lambda }^{H}( \mathbf{x}) \mathbf{f}_{B\Lambda }(
\mathbf{x}) $ and $\mathbf{f}_{B\Lambda }( \mathbf{x})
\triangleq \mathbf{g}_{B\Lambda }( \mathbf{x}) \mathbf{H}(
\mathbf{x}) .$\

\subsection{Secrecy Performance Metrics}

The secrecy capacity is the performance metric we adopt in this paper to
evaluate the performance of the overall system, which can be formulated for
securely decoding $u_{g}$ in the presence of $T_{k}$ as
\begin{equation}
C_{s}^{(g,k)}\left( \left\{ \mathbf{V}_{g}\right\} _{g=1}^{G},\mathbf{W},%
\mathbf{x}\right) =\left[
\begin{array}{c}
C^{(g)}_{U_{g}}\left( \left\{ \mathbf{V}_{g}\right\} _{g=1}^{G},\mathbf{W,x}%
\right) \\
-C_{T_{k}}^{(g)}\left( \left\{ \mathbf{V}_{g}\right\} _{g=1}^{G},\mathbf{W%
\mathbf{,}x}\right)%
\end{array}%
\right] ^{+},  \label{cs}
\end{equation}%
where $\left[ t\right] ^{+}\triangleq \max \left( 0,t\right) $, while
\begin{equation}
C_{\Lambda}^{(g)}\left( \left\{ \mathbf{V}_{g}\right\} _{g=1}^{G},\mathbf{W,x%
}\right) =\log _{2}\left( 1+\gamma^{(g)} _{\Lambda}\left( \left\{ \mathbf{V}%
_{g}\right\} _{g=1}^{G},\mathbf{W,x}\right) \right)  \label{ccdlgeneral}
\end{equation}%
is the channel capacity of the link to node $\Lambda$, computed using (\ref%
{snruser}). Due to the consideration of multiple legitimate users and
malicious eavesdroppers, the equivalent system's SC is defined as the
worst-case SC, i.e.,
\begin{equation}
C_{s}\left( \left\{ \mathbf{V}_{g}\right\} _{g=1}^{G},\mathbf{W},\mathbf{x}%
\right) =\min_{\substack{ g=1,\ldots ,G  \\ k=1,\ldots ,K}}%
C_{s}^{(g,k)}\left( \left\{ \mathbf{V}_{g}\right\} _{g=1}^{G},\mathbf{W},%
\mathbf{x}\right).  \label{Csdl}
\end{equation}
\vspace*{-1cm}
\subsection{Sensing Performance Metrics}

The considered system desires ensuring monostatic radar sensing to detect
the presence of the $K$ malicious targets. By virtue of the flexible channel
response adjustment through PA location fine-tuning, the ISAC\ transceiver
aims at establishing a robust beamsteering through the set of waveguides and
antennas to illuminate the $K$ targets with a maximal electromagnetic
power.\ The higher the latter, the greater the echo signal reflected back to
the ISAC\ receiver, maximizing the target detection probability. Thus, the
sensing performance is quantified by the illumination power, defined as
\begin{equation}
Q_{s}^{(k)}\left( \mathbf{R}_{s}\mathbf{,x}\right) ={\sigma
_{RCS}^{(k)}4\pi }\mathrm{Tr}\left[ \mathbf{F}_{BT_{k}}\left(
\mathbf{x}\right) \mathbf{R}_{s}\right] /G^{(T_{k})} ,  \label{senspw}
\end{equation}%
where $\mathbf{R}_{s}\triangleq \sum_{g=1}^{G}\mathbf{V}_{g}+\mathbf{W%
}$. In (\ref{senspw}), $Q_{s}^{(k)}(\mathbf{R}_s, \mathbf{x}) $
indicates the level of electromagnetic power illuminating $T_{k}$, where $%
\sigma _{RCS}^{(k)}$ is the radar cross section (RCS)\ of $T_{k}$.

Another metric for evaluating the sensing performance is the illumination
power pattern, expressed as
\begin{equation}
Q_{s}\left( \mathbf{R}_{s},\mathbf{x},x_{0},y_{0},z_{0}\right) =\overline{%
\sigma }_{RCS}4\pi \mathrm{Tr}\left[ \mathbf{F}_{0}\left( \mathbf{x}%
,x_{0},y_{0},z_{0}\right) \mathbf{R}_{s}\right] ,  \label{illumpattern}
\end{equation}%
where $\mathbf{F}_{0}( \mathbf{x},x_{0},y_{0},z_{0}) \triangleq
\mathbf{f}_{0}^H( \mathbf{x},x_{0},y_{0},z_{0}) \mathbf{f}%
_{0}( \mathbf{x},x_{0},y_{0},z_{0}) $, $\mathbf{f}_{0}(
\mathbf{x},x_{0},y_{0},z_{0})=\mathbf{g}_0(\mathbf{x},x_0,y_0,z_0
) \mathbf{H}( \mathbf{x} )$, with $\mathbf{g}_0(%
\mathbf{x},x_0,y_0,z_0 )$ can be computed from (\ref{glos}) and (\ref%
{hx}) by setting $G^{(\Lambda )}=1$ in the FSPL\ term, computed using (\ref%
{fspl}), while the distance $d_{B_{n,m}\Lambda }$ is substituted by $%
d_{B_{n,m},P_{0}}$, evaluated from (\ref{distance}) by substituting $\left(
x_{\Lambda },y_{\Lambda },z_{\Lambda }\right) $ by $\left(
x_{0},y_{0},z_{0}\right) $. Also, $\overline{\sigma }_{RCS}$ is the average
RCS across the $K$ targets. Note that $Q_{s}\left( \mathbf{R}_{s},\mathbf{x}%
,x_{0},y_{0},z_{0}\right) $ is similar to the sensing beampattern metric,
measuring the transmit signal power's angular directivity over the BS's
angular look direction, while $Q_{s}\left( \mathbf{R}_{s},\mathbf{x}%
,x_{0},y_{0},z_{0}\right) $ assesses the transmit power illumination level
at each point $P_{0}\left( x_{0},y_{0},z_{0}\right) $ served by the array of
PAs, which effectively evaluates the level of power reaching the locations
of interest, i.e., targets' locations.

\section{Secure PA-ISAC\ Optimization}
\subsection{Optimization Problem Formulation}
The optimization problem under consideration aims at maximizing the sensing
performance, subject to secrecy constraints. The PA-ISAC system utilizes
pinching beamforming by leveraging the set of PAs of the various waveguides.
The set of PAs offers flexibility in adjusting the\ wireless and
in-waveguide channel response, i.e., (\ref{glos}) and (\ref{hx}), whereas
the beamforming and AN\ covariance matrices can further enhance signal
beamsteering. In the considered design, the sensing performance is
prioritized. Thus, the optimization framework consists of optimizing the PA
locations, the transmit signal beamforming and AN\ covariance matrices in
order to maximize the sensing illumination power per target, subject to
total power and secrecy constraints. This is formulated as
\begin{subequations}
\label{P1}
\begin{align}
\mathcal{P}1& :\max_{\left\{ \mathbf{V}_{g}\right\} _{g=1}^{G},\mathbf{W,x}%
}\min_{k=1,\ldots ,K}Q_{s}^{(k)}\left( \mathbf{R}_s,\mathbf{x}\right) \\
\text{s.t.}\ (\mathrm{C1})& :C_{s}\left( \left\{ \mathbf{V}_{g}\right\}
_{g=1}^{G},\mathbf{W,x}\right) \geq C_{s,\mathrm{lb}}  \label{C1a} \\
(\mathrm{C2})& :\mathrm{rank}\left( \mathbf{V}_{g}\right) =1,\forall g
\label{C2a} \\
(\mathrm{C3})& :\mathbf{V}_{g}\succeq \mathbf{0},\forall g,\mathbf{W}\succeq
\mathbf{0}  \label{C3a} \\
(\mathrm{C4})& :\sum\limits_{g=1}^{G}\mathrm{Tr}\left[ \mathbf{V}_{g}\right]
+\mathrm{Tr}\left[ \mathbf{W}\right] \leq P_{\max }  \label{C4a} \\
(\mathrm{C5})& :x_{n,m}-x_{n,m-1}\geq \Delta x,\forall n,\forall m\geq 2
\label{C5a} \\
(\mathrm{C6})& :x_{n,m}\geq x_{0},x_{n,m}\leq x_{\max }  \label{C6a}
\end{align}
\end{subequations}
where $(\mathrm{C1})$ is the minimal secrecy requirement constraint
with $C_{s,\mathrm{lb}}$ representing the minimal network's SC\ requirement,
while $(\mathrm{C2})$ and $(\mathrm{C3})$ define the rank-one property of $%
\left\{ \mathbf{V}_{g}\right\} _{g=1}^{G}$ and the positive semidefinitness
property of the latter and $\mathbf{W}$, where\textbf{\ }$\mathbf{A}\succeq 0
$ denotes that $\mathbf{A}$ is a PSD matrix. $(\mathrm{C4})$ defines the
total power budget constraint, and $(\mathrm{C5})$-$(\mathrm{C6})$ refer to
the constraints on each PA's location, where $\Delta x$ is the minimal
inter-PA separation, whereas $x_0$ and $x_{\max}$ define the interval for
each PA's position. $\mathcal{P}1$ is challenging to handle due to the coupling
between the PAs locations-dependent channel matrix $\mathbf{F}_{B\Lambda
}\left( \mathbf{x}\right) $ and $\{ \mathbf{V}_{g}\} _{g=1}^{G},%
\mathbf{W}$, as noted from (\ref{snruser}) and (\ref{senspw}). Also,
for given $\{ \mathbf{V}_{g}\} _{g=1}^{G},\mathbf{W}$, another
hurdle to optimize $\mathbf{x}$ is the non-convexity of $Q_{s}^{(k)}(
\{ \mathbf{V}_{g}\} _{g=1}^{G},\mathbf{W,x}) $ and $%
C_{s}( \{ \mathbf{V}_{g}\} _{g=1}^{G},\mathbf{W,x}) $
in $\mathbf{x}$, as noted from (\ref{snruser}) and (\ref{senspw}). This is
caused essentially by the involvement of $\mathbf{x}$'s elements $(x_{n,m})$
in the complex exponentials in (\ref{glos}) via the distances $%
d_{B_{n,m}U_{g}}$, $d_{B_{n,m}T_{k}}\left( \forall g,k\right) $ and also in (%
\ref{hx}). It is worth mentioning that previous work aimed to solve similar
problems in PASS\ using alternating optimization, such as in \cite%
{lit2,isaclit4}, in which $\mathbf{x}$ was optimized over two steps, by
first solving for a slack channel matrix $\mathbf{Z}$ maximizing the sensing
or reliability performance, followed by performing an iterative element-wise
line search to sequentially optimize (i.e., one PA location at once) the PAs
locations producing the nearest channel response to $\mathbf{Z}$.
Nonetheless, the schemes in \cite{lit2,isaclit4} did not consider secrecy
constraints and the one in \cite{isaclit4} was limited to a single
user/target case. In addition, note that the line search-based
solution for $\mathbf{x}$ in the above work may produce a solution violating the constraints
of $\mathcal{P}1$. Motivated by this, a novel algorithm for optimizing the PA
locations is presented in the sequel.

\subsection{PA Positions Optimization}

The proposed PA positions optimization scheme is based on\ positioning the
set of PAs in each waveguide closer to the users and targets to maximize
their respective channel gains. Without loss of generality, it is assumed
that $M_{t}>G$ and $M_{t}>K$. For each waveguide, the proposed approach
starts by placing the first $G\ $PAs aligned with the $G$ users, i.e., $%
x_{n,g}=x_{U_{g}}$, $g=1,\ldots ,G$, $\forall n$. Then, the set of targets
are evaluated and ranked in an ascending order in terms of their channel
magnitudes utilizing only the currently positioned PAs, i.e.,
\begin{equation}
P_{k}^{(n)}=\left\Vert \overline{\mathbf{g}}_{BT_{k}}^{(n)}\left( \overline{%
\mathbf{x}}^{(n)}\right) \mathbf{H}\left( \overline{\mathbf{x}}^{(n)}\right)
\right\Vert ^{2}  \label{magnitude}
\end{equation}%
where $P_{k}^{(n)}$ is the evaluated channel gain for $T_{k}$ while
optimizing the PA locations at the $n$th waveguide, with $\overline{\mathbf{x%
}}^{(n)}\in
\mathbb{R}
^{1\times \left( \left( n-1\right) M_{t}+G\right) }$ is a vector consisting
of the optimized $x$-axis coordinates of the $(n-1)M_t+G$ PAs already
positioned. Also,%
\begin{equation}
\overline{\mathbf{g}}_{BT_{k}}^{(n)}\left( \overline{\mathbf{x}}%
^{(n)}\right) \triangleq \left[
\begin{array}{c}
\mathbf{g}_{B_{1}T_{k}}\left( \left[\overline{\mathbf{x}}^{(n)} \right]%
_{1:M_{t}}\right) ,\ldots , \\
\left[ \mathbf{g}_{B_{n}T_k }\left( \left[ \overline{\mathbf{x}}^{(n)}\right]
_{\left( n-1\right) M_{t}+1:\left( n-1\right) M_{t}+G}\right) \right] _{1:G}
\\
,\mathbf{0}_{1\times \left( \left( N-n+1\right) M_{t}-G\right) }%
\end{array}%
\right],
\end{equation}%
with $\overline{\mathbf{g}}_{BT_{k}}^{(n)}( \overline{\mathbf{x}}%
^{(n)}) \in
\mathbb{C}
^{1\times \left( N M_{t} \right) }$ being a zero-padded vector with $\left(
n-1\right) M_{t}+G$ channel response elements of the positioned PAs with $%
\left[ \mathbf{q}\right] _{a:b}$ denotes the portion of a vector $\mathbf{q}$
between the indices $a$ and $b$. Then, the $K$ targets are sorted by their
channel magnitudes, i.e., $P_{i_{K}}^{(n)}\geq \ldots \geq P_{i_{1}}^{(n)}$,
with $i_{k}$ defining the target's index with the $k$th lowest channel
magnitude. Afterwards, the remaining $M_{t}-G$ PAs in the $n$th waveguide
are set in the current order of increasing channel magnitude of the $K$
targets, i.e., starting from the target with the weakest channel magnitude ($%
T_{i_1}$). A PA is aligned with a target as follows: $x_{n,G+k}=x_{T_{i_{k}}}
$, $k=1,\ldots ,K$, assuming $|
x_{T_{i_{k}}}-x_{T_{i_{k-1}}}| \geq \Delta x,\forall k\geq 2$.
Note that in the case when $K>M_{t}-G$, a proportion of $K-M_{t}+G$ targets
with the highest channel magnitude will not benefit from the optimized
antenna placement according to their locations, as the proposed scheme
focuses on enhancing the channel magnitude of the $M_{t}-G$ targets with the
lowest channel magnitude. On the other hand, observe that when $K<M_{t}-G$
and $M_{t}-G\neq qK (q \in \mathbb{N}^{\ast})$, some targets can have more
than one PA aligned to its $x$-axis coordinate per the above-mentioned
placement rule, where after placing the first $K$ PAs out of the $M_{t}-G$
remaining ones, the considered scheme starts over again from $T_{i_{1}}$
until completing positioning all the $n$th waveguide's PAs . In this case,
it can be noted that co-locating two or more antennas at $x_{T_{i_{k}}}$
violates (\ref{C5a}).\ Therefore, in this scenario, the proposed scheme
first checks when placing the $\left( m+G\right) $th PA, if any of the
already-positioned PAs is close to $x_{T_{i_{\left( m+G-1\right) \mathrm{mod}%
K+1}}}$ by less than $\Delta x$. If the latter condition is satisfied, the
scheme performs a line search over the waveguide to find the location
minimizing the distance to the current target in consideration, i.e., $%
T_{i_{\left( m+G-1\right) \mathrm{mod}K+1}}$ while fulfilling \eqref{C5a}
with already-positioned PAs in the same waveguide, i.e.,
\begin{align}
x_{n,m+G} &=\min_{x}\left( \left\vert x-x_{T_{i_{\left( m+G-1\right) \mathrm{%
mod}K+1}}}\right\vert \right), m=1,\ldots,M_t-G  \notag \\
\text{s.t. } &:\text{ }\left\vert x-x_{n,p}\right\vert \geq \Delta
x\left(\forall p<m+G\right) \text{ } \& \text{ } x\in \left[ x_{0},x_{\max }%
\right] .  \label{linesearch}
\end{align}
The aforementioned process is performed identically for the remaining
waveguides.









\subsection{Beamforming and AN Optimization}

For an optimized PA positions vector $\mathbf{x}^{\mathrm{(opt)}}$ using the proposed scheme in the previous subsection, the
optimization of $\{ \mathbf{V}_{g}\} _{g=1}^{G},\mathbf{W}$ is
carried out by dropping $\mathbf{x}$ from the control variables of \eqref{P1}%
. Observe that $\mathcal{P}1$ is challenging to optimize due to (i) the complex
form of the objective function, i.e., max-min fairness problem, (ii) the
non-convex SC expression in (\ref{C1a}) in terms of $\{ \mathbf{V}%
_{g}\} _{g=1}^{G},\mathbf{W}$, and the non-convex rank-one constraint
in (\ref{C2a}).\ To overcome this hurdle, the following alternative
representation is considered
\begin{subequations}
\label{P2}
\begin{align}
\mathcal{P}2& :\max_{\left\{ \mathbf{V}_{g}\right\} _{g=1}^{G},\mathbf{W,}\theta }\rho
\\
\text{s.t.}\ (\mathrm{C1})& :\gamma _{U_{g}}^{(g)}\left( \left\{ \mathbf{V}%
_{g}\right\} _{g=1}^{G},\mathbf{W,x}^{\mathrm{(opt)}}\right) \geq \gamma _{U,%
\mathrm{lb}},\forall g  \label{C1b} \\
(\mathrm{C2})& :\gamma _{T_{k}}^{(g)}\left( \left\{ \mathbf{V}_{g}\right\}
_{g=1}^{G},\mathbf{W,x}^{\mathrm{(opt)}}\right) \leq \gamma _{E,\mathrm{ub}%
},\forall g,k  \label{C2b} \\
(\mathrm{C3})& :Q_{s}^{(k)}\left( \mathbf{R}_{s},\mathbf{x}^{\mathrm{(opt)}%
}\right) \geq \rho  \label{C3b} \\
& (\ref{C2a})-(\ref{C6a})
\end{align}
In (\ref{P2}), the SC\ constraint in (\ref{C1a}) was replaced by $\mathrm{C1}
$ and $\mathrm{C2}$ in (\ref{C1b}) and (\ref{C2b}). The latter constraints
can equivalently maintain a target minimal SC\ of the network by imposing a
maximal received SINR\ at each malicious target $\gamma _{E,\mathrm{ub}}$
while preserving a minimal SINR $\gamma _{U,\mathrm{lb}}$ for each user for
reliable signal decoding. Note from (\ref{snruser}), (\ref{C1b}) and (\ref%
{C2b}) that $\mathrm{C1}$ and $\mathrm{C2}$ of (\ref{P2}) are convex. In
addition, $\theta $ is involved as a slack variable to tackle the complex
objective function form in (\ref{P1}).\ Herein, $\rho $ is linked with the
sensing illumination power of each target via $\mathrm{C3}$ in (\ref{C3b}),
which aims at maximizing the per-target illumination power. By relaxing the
rank-one constraint in (\ref{C2a}), $\mathcal{P}2$ becomes a convex semidefinite
program, which can be solved by any convex optimization tool,
e.g., CVX. Then, an eigenvalue decomposition-based approach is used to
transform $\left\{ \mathbf{V}_{g}\right\}_{g=1}^G$ into rank-$1$ solutions
by expressing them in terms of their principal eigenvectors \cite{bazzi}.

\section{Numerical Results}

This section presents numerical examples through which the performance of
the proposed scheme is evaluated. Table \ref{sysparam} indicates the values
of the various system parameters used in the simulations. The wavelength is
linked to carrier frequency $f_{c}$ as $\lambda =c/f_{c}$, where $c$ is the
speed of light in the vacuum. Furthermore, it is considered that all nodes
and targets are positioned at the ground level, i.e., $%
z_{U_{g}}=z_{T_{k}}=0,\forall t,k$. For the set of users and targets, the
various nodes are distributed equidistantly over their respective angular
and radius intervals given in Table \ref{sysparam}, where $d_{O\Lambda }$ is
the distance from the origin $\left( x_{O}=0,y_{O}=0\right) $ to $\Lambda $.
In addition, $\varphi_{O\Lambda}$ denotes the azimuth relative angle of node
$\Lambda$ with respect to the origin $O$.
\begin{table}[h]
\captionsetup{font=footnotesize}
\caption{System Parameters Values} \label{sysparam}\centering%
\begin{tabular}{|c|c||c|c|}
\hline\hline
Parameter & Value/Interval & Parameter & Value/Interval \\ \hline\hline
$f_{c}$ & $15$ GHz & $G^{\left(B_{n,m}\right)} $ & $30$ dB \\ \hline
$G^{\left(\Lambda\right)}$ & $10$ dBi & $d_{OU_{g}} $ & $\left[ 40, 50 %
\right]$ m \\ \hline
$d_{OT_{k}} $ & $\left[ 20, 40 \right]$ m & $\varphi _{OU_{g}}$ & $%
[-50,-40]^{\circ}$ \\ \hline
$\varphi _{OT_{k}} $ & $30^{\circ}$ & $y_{{n}}$ & $15+5n$ m \\ \hline
$z_{n}$ $(\forall n)$ & $10$ m & $\sigma _{\Lambda}^{2} $ & $-110$ dB \\ \hline
$G$ & $2$ & $K$ & $3$ \\ \hline
$N$ & $8$ & $M_t$ & $6$ \\ \hline
$P_{\max}$ & $20$ dB & $\gamma_{E,\mathrm{ub}}$ & $3$ dB \\ \hline
$x_0$ & $0$ m & $x_{\max}$ & $60$ m \\ \hline
\end{tabular}%
\end{table}
\begin{figure}[h]
\vspace*{-.2cm}\centering  
\includegraphics[scale=.41]{
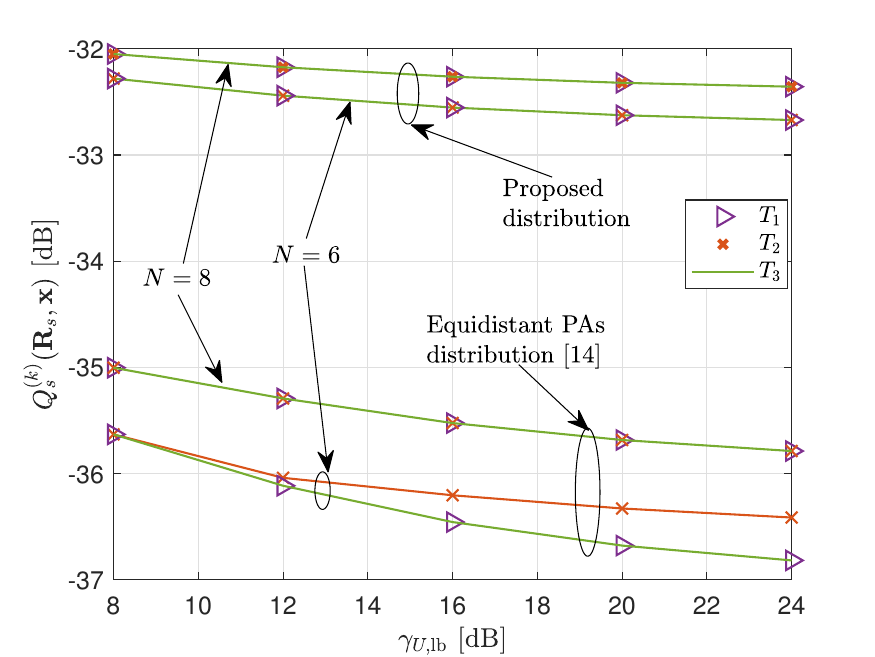}
\captionsetup{font=footnotesize}
\caption{Illumination power of the proposed scheme vs. $\protect\gamma _{U,%
\mathrm{lb}}$ compared with an equidistant PAs deployment \cite{benchref}.}
\label{fig3}
\end{figure}
\begin{figure}[h]
\vspace*{-.4cm}
\centering  
\includegraphics[scale=.41]{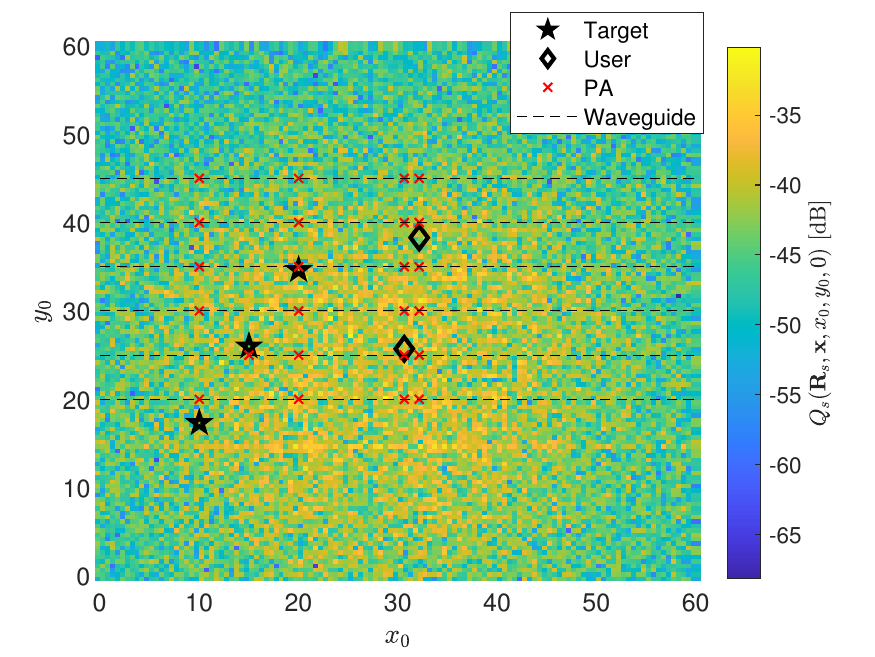}
\captionsetup{font=footnotesize}
\caption{Illumination power pattern over the considered communication and sensing area, along with the optimized PAs positions.}\label{fig5}
\end{figure}

In Fig. \ref{fig3}, the proposed scheme's performance in terms of the
illumination power is shown versus $\gamma _{U,\mathrm{lb}}$ and compared
against a baseline equidistant PAs positioning \cite{benchref}. The latter
approach positions the set of $M_{t}$ PAs along each waveguide equidistantly
with the first and last PAs at $x_{0}$ and $x_{\max }$, respectively.
Observe that the proposed PAs positioning approach yields a higher target
illumination power compared to the aforementioned baseline scheme, where $4$%
-dB sensing gain is observed at $\gamma _{U,\mathrm{lb}}=24$ dB with $N=6$
compared to the baseline scheme. Observe that per the proposed algorithm,
only $M_{t}=G=2$ PAs from each waveguide are positioned close to the
corresponding users, whereas the remaining $M_{t}-G=6$ PAs, representing a
higher proportion, are placed in such a way to enhance the channel condition
of the sensed targets. Also, observe the slight illumination power increase
by increasing the network secrecy requirement in terms of $\gamma_{U,\mathrm{%
lb}}$, showing an existing secrecy-sensing trade-off. In Fig. \ref{fig5},
the illumination power pattern over the considered geographical area is
shown, evaluated using \eqref{illumpattern} with $N=6$ and $M_t=4$. The obtained results indicate a
relatively large spot in the center of the area with scattered points of
high illumination power.\ The sensed targets and users are positioned in
relatively highly-illuminated positions, with $Q_{s}^{(1)}=Q_{s}^{(2)}=-39$
dB while $Q_{s}^{(3)}=-41.5$ dB.

In Fig. \ref{fig6}, the proposed scheme's sensing power is presented and
compared against the baseline secure ISAC\ scheme in \cite{bazzi}. The
latter consists of a dual-functional radar communication BS with a ULA of
half-wavelength-spaced antennas, whereas the scheme aims at minimizing the
total transmit power subject to the secrecy constraints in \eqref{C2b} and %
\eqref{C3b} and to a sensing constraint. The latter is defined by a maximal
sidelobe-to-mainlobe ratio, chosen to $-20$ dB in our evaluation. For the
sake of fairness, UL\ transmission is not considered in the baseline scheme,
and $P_{\max }$ for the proposed framework is set as the optimized power by
the benchmark one. Also, we set $\gamma_{E,\mathrm{ub}}=-6$ dB and $N=6$.
Note that the proposed and baseline schemes yield an increasing sensing
power with the increase of $\gamma _{U,\mathrm{lb}}$. This is due to the
fact that a higher $\gamma _{U,\mathrm{lb}}$ requires an increased signal
and AN\ power for legitimate signal and AN beamforming, resulting in
increased illumination power. Furthermore, the proposed approach yields a
high sensing illumination power compared with the baseline scheme, exceeding
$30$ dB.
\begin{figure}[h]
\centering  
\vspace*{-.2cm}\includegraphics[scale=.41]{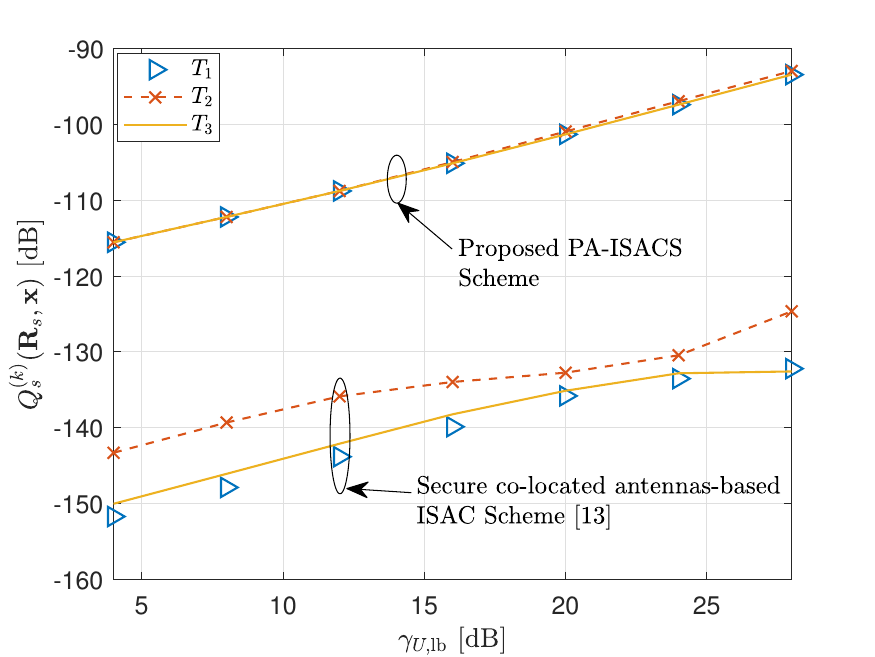}
\captionsetup{font=footnotesize}
\caption{Illumination power of the proposed scheme compared with the secure
ISAC scheme of \protect\cite{bazzi}.} \label{fig6}
\end{figure}
\vspace*{-.2cm}
\section{Conclusions}

In this letter, we analyzed a robust and secure PA-ISAC system. The proposed
approach invoked pinching beamforming to securely beamsteer information
signals to various DL users in the presence of malicious targets, while
maximizing the sensing signal power for the latter. A novel algorithm for
obtaining suboptimal PA positions was proposed. Those locations were used to design
optimal beamforming and AN\ covariance matrices, using semidefinite
programming methods. The obtained results showed a notable enhancement in
terms of sensing illumination power compared to the baseline secure ISAC
scheme with traditional half-wavelength-spaced ULA.

\bibliographystyle{IEEEtran}
\bibliography{refs}
\end{subequations}

\end{document}